\documentclass[letterpaper,conference]{IEEEtran}
\usepackage{color}
\usepackage{epsfig}
\usepackage{amsmath}
\usepackage{amsfonts}
\usepackage{amsthm}
\usepackage{amssymb}
\usepackage{graphics}
\usepackage{graphicx}
\usepackage{float}
\usepackage{algorithm}
\usepackage{algorithmic}
\usepackage{svg}
\usepackage {epstopdf}

\title{\LARGE \bf A Simple Sign-Bit Probabilistic Shaping~Scheme }

\author{Vincent Corlay and Nicolas Gresset\\
\small{Mitsubishi Electric R\&D Centre Europe, Rennes, France. E-mail: \{v.corlay, n.gresset\}@fr.merce.mee.com.} }

\begin{document}

\maketitle

\begin{abstract}
We propose a new shaping scheme for the Gaussian channel whose complexity is approximately half the one of a binary distribution matcher (DM).
The result is obtained as follows:
We first show that most of the shaping gain can be obtained via a simplified version of sign-bit shaping, which uses only two non-uniform binary sources. This is achieved by considering a stepwise Maxwell-Boltzmann-like distribution of the symbols. 
One of the two binary sources has a parameter $p$ close to 0. Hence, we then describe a binary DM which explicitly takes advantage of this aspect and has a negligible complexity. Since the two binary sources are used alternately with equal probability, the complexity of the proposed shaping scheme is half the one of the second binary DM.

\end{abstract}



\vspace{-5mm}
\section{Introduction}
\label{sec_intro}

The channel capacity characterizes the highest information rate that can be achieved for a fixed average transmit power while maintaining a small error probability. 
For the Gaussian channel, the capacity cannot be reached if each symbol of commonly used constellations, such as the amplitude-shift keying (ASK), is transmitted with equal probability. 
As a result, the transmitter should process the data such that the symbols of the constellation are transmitted with a probability which enables to approach the capacity. 
This operation is called shaping. More precisely, it is called probabilistic shaping, to be opposed with geometric shaping\footnote{The main idea of geometric shaping is to change the position of the symbols in the constellation without changing the probability distribution. 
}\cite{Boutros2018}.

\vspace{-2mm}
\section{Relevant Literature and main contributions}

\vspace{-1mm}
\subsection{Existing probabilistic shaping methods}

With the popular probabilistic amplitude shaping scheme (PAS)~\cite{Bocherer2015}, a symbolwise DM is considered. 
However, a new paradigm for shaping has recently emerged: 
It consists in using several binary DMs, instead of one symbolwise DM.
There are two main justifications for this approach. 
The first is the computational complexity: As stated in \cite{Bohnke2020B}, the  complexity of a non-binary DM grows linearly with the size of the constellation (exponentially with the number of bit levels).
The second is the latency: As reported in \cite{Pikus2017}, bitwise DMs allow for parallel processing of input bits unlike a symbol-wise DM. 

These bitwise approaches can be classified into two categories:
\begin{itemize}
\item \textbf{Method 1}: Independent bit-level distribution matching   \cite{Pikus2017}\cite{Iscan2018}\cite{Steiner2018}\cite{Miller2019}.
\item \textbf{Method 2}: Conditional bit-level distribution matching   \cite{Matsumine2019}\cite{Bohnke2020}\cite{Bohnke2020B}\cite{Forney1992}.
\end{itemize}

We illustrate the two methods with a $M$=8-ASK and natural labelling. 
The symbols of this constellation, with $M=2^m$, are
\small
\vspace{-2mm}
\begin{align}
\mathcal{X}=\{-2^m-1,..,-3,-1,+1,+3,…,+2^m-1\},
\end{align}
\normalsize
where each symbol can be represented by a sequence $b_1 b_2 ...b_m$ of $m=\log_2 M$ bits.

With Method 1, each bit level is assigned with an independent probability $p(b_i=0)=p_i$.
For instance, the bit sequence $0,0,0$ has probability $p_1 p_2 p_3$, and the bit sequence $0,1,1$ probability $p_1 (1-p_2) (1-p_3)$. 

With Method 2, the probability of a bit at a given level depends on the value of the bits at the previous levels.
For the first level, we have $p(b_1=0)=p_1$. For the second level, $p(b_2=0|b_1=0)=p_2$ and $p(b_2=0|b_1=1)=p_3$.
For the last level, $p(b_3=0|b_1=0,b_2=0)=p_4$, $p(b_3=0|b_1=1,b_2=0)=p_5$, $p(b_3=0|b_1=0,b_2=1)=p_5$,  $p(b_3=0|b_1=1,b_2=1)=p_6$.
Consequently, the bit sequence $0,1,1$ has probability $p_1 (1-p_2) (1-p_5)$.


\vspace{-2mm}
\subsection{Optimization of the parameters}
\vspace{-1mm}

For methods 1 and 2, the parameters $p_i$ should be chosen to optimize the performance. For the Gaussian channel, they can be established such that the resulting distribution of the symbols approaches the Maxwell-Boltzmann (MB) distribution, as done in \cite{Pikus2017}\cite{Bohnke2020B}. More generally, they should be chosen to maximize the mutual information (MI) between the input of the channel $X$ and the output $Y$

\vspace{-5mm}
\small
\begin{align}
\label{MI_eq}
\{p_i^*\}=  \underset{\{p_i\}, E[X^2] \leq P }{\text{arg max} }   I(X;Y),
\vspace{-3mm}
\end{align}
\normalsize
where $P$ is the maximum average power. 


When optimizing the parameters $p_i$, many authors noticed that several bit levels remain uniform and only a subset of the bit levels needs to be shaped. One can even fix the value of some parameters, while optimizing the remaining ones, and observe how it affects the MI.

For instance, if only one bit level is used with method~2 and natural labelling, this is called sign-bit shaping \cite{Forney1992}, applied in \cite{Wachsmann1999} and studied recently in \cite{Bohnke2020}. 
In the scope of probabilistic shaping, standard sign-bit shaping requires $M/2$ distinct non-uniform binary sources.
Alternatively, instead of targetting a specific probability distribution, the metric for a sign-bit shaping decoder can be the following: Chose the shaping bits such that the resulting shaped sequence of symbols has minimum energy as done in \cite{Forney1992}\cite{Wachsmann1999} and more recently in~\cite{Matsumine2021}.

More generally, several bit levels, $s<m$, can be jointly shaped while others remain unshaped, as in \cite{Forney1992} with trellis shaping.
These $s$ levels can be shaped with a symbolwise DM working on a reduced number of symbols ($2^s$ instead of $2^m$), as done in \cite{Gultekin2019} with enumerative sphere shaping, and in \cite{Yoshida2019B} with hierarchical distribution matching and constant composition distribution matching. 

Consequently, many authors shape only a subset of the bit levels: 3 in \cite{Steiner2018} (with method 1, which yields 3 distinct $p_i$), 2 in \cite{Iscan2018} (with method 1, which yields 2 distinct $p_i$), 2 in \cite{Bohnke2020B}  (with method 2 and Gray labelling, which yields $M/2+M/4$ distinct $p_i$), and 1 in \cite{Bohnke2020} with sign-bit shaping (which yields $M/2$ distinct $p_i$).

In \cite{Gultekin2019}, the authors noticed that a stepwise MB distribution, where several adjacent symbols have the same probabilities, yields very satisfactory performance. They use this distribution to reduce to 2 the number of bit levels considered for shaping via a symbolwise DM. 
The authors of \cite{Iscan2018} made a similar observation.
\vspace{-2mm}
\subsection{Binary distribution matcher}
\label{sec_distri_match}

Let $H(p)$ denote the binary entropy with parameter $p$.
An ideal binary DM takes an equiprobable sequence $v$ of size $n H(p)$ as input and outputs a sequence $c$ of size $n$ where each bit is i.i.d. with a Bernoulli($p$) distribution. 

A first approach, to realize a binary DM, is to design a non-linear code of size $\approx 2^{n H(p)}$ where all codewords have a Hamming weight equal to\footnote{Here, for simplicity, $p=p(b=1)$.} $n p$. Then, any sequence $v$ should be mapped to one of these codewords. 
Algorithms to index sequences of length $n$ and weight $w$ are presented in \cite{Schalkwijk1972}\cite{Ramabadran1990}. The algorithm of  \cite{Schalkwijk1972} is discussed in Section~\ref{perf_with_DM} and the one of \cite{Ramabadran1990} (arithmetic coding) is used in \cite{Pikus2017}\cite{Steiner2018}.

Another approach is to select codewords of a larger linear code that have a Hamming weight equal to $np$.
This linear code can for instance be a polar code \cite{Bohnke2019B} as done in \cite{Matsumine2019}\cite{Bohnke2020}\cite{Bohnke2020B}. The principle is the following: Given a target probability $p$, a set of frozen indices is determined for the BSC$_p$ channel. 
The values of the frozen bits are dynamically set to $v$. 
The polar decoder assumes that the all 0 sequence is received on a BSC$_p$ channel. 
Since this channel introduces on average $np$ errors, the decoded sequence $c$ should have a Hamming weight of $np$.
If soft-input decoding is used, the inputs of the decoder are the LLR log$((1-p)/p)$.
The decoded bits on the information indices are the shaping bits. 

If multi-level coding is considered, the binary DMs should be combined with binary error correcting codes, as in \cite{Matsumine2019}\cite{Bohnke2020B}. 
This is not trivial to implement:  The binary DM should be placed before the channel code, but the sequence after the channel coding should have the desired distribution. We refer to this topic as \textbf{problem 1}.
Note however that the design of polar codes for joint shaping and channel coding has been studied in several publications as e.g., \cite{Honda2013}\cite{Liu2019}. 
The advantage of the PAS scheme is that it avoids the concatenation of a DM and a channel code.
We shall see that sign-bit shaping also avoids this issue.

Besides, if method 2 is used, the bits should have varying probabilities within one level depending on the value of the bits at the previous levels. 
This bitwise approach to fit a probability distribution requires using several binary DMs in parallel. 
For a given packet size, this implies reducing the length of the output sequence of each DM, which increases the finite-length rate loss. 
We refer to this issue as \textbf{problem~2}.
Alternatively, the authors in \cite{Bohnke2020}\cite{Bohnke2020B} propose to set the input LLR (of one binary DM polar decoder) corresponding to each bit according to the desired distribution. 
Nevertheless, the efficiency of this approach is not clear: How should the polar code be designed?  
Moreover, at a given bit location, the input LRR changes between two codewords due to the conditioning on the bit values on the previous levels.
In \cite{Bohnke2020}\cite{Bohnke2020B} standard 5G NR polar codes based on a universal polar sequence are used for both shaping and coding.
Based on the results presented, this heuristic seems satisfactory and can therefore be considered as an interesting alternative to the proposed shaping scheme.
However, the same authors state in \cite{Bohnke2020B} that ``in order to reduce the complexity and facilitate a parallel implementation, independent distribution matchers may be applied to different subsequences".  

Finally, the value of $p$ impacts the complexity of most algorithms utilized to implement a binary DM.
In Section~\ref{perf_with_DM}, we explain that the complexity of the algorithm in \cite{Schalkwijk1972} is linear in $p$ and therefore negligible if $p$ is close to 0.
Similarly, the complexity of successive-cancellation decoding of a polar code highly depends on the number of frozen bits:  \cite{Alamdar2011}\cite{Ali2016} show that the decoding tree can be pruned if all leaves of a sub-tree are frozen bits, which improves both the latency and the complexity.
Unfortunately, method 1 requires ``balanced" $p_i$. As an example with the 8-ASK, method 1 requires two DMs with relatively high $p_i$: $p_1=0.33$ and $p_2=0.25$ (see Fig.~4 in \cite{Pikus2017}). We refer to this issue as \textbf{problem~3}.

\vspace{-2mm}
\subsection{Main contributions}
\vspace{-1mm}

We simplify sign-bit shaping by limiting the number of distinct parameters~$p_i$. 
This leads to a stepwise MB-like distribution, similar to the one considered\footnote{We exploit this observation in a different manner from \cite{Gultekin2019}: 
We consider a bitwise probabilistic approach via method 2 whereas \cite{Gultekin2019} uses a symbolwise DM to select sequences of low energy.} 
in \cite{Gultekin2019}.
This heavily quantized MB-like distribution does not significantly alter the performance.
As a result,  most of the shaping gain can be obtained via the use of only two non-uniform binary sources to shape only one bit level. 
Moreover, the parameter $p$ of one of  the two binary sources is very close to 0.
This is interesting with respect to problem~3.
We describe a binary DM which explicitly takes advantage of this aspect and has negligible complexity.
Consequently, most of the complexity lies in the second DM.
Since the two binary DMs are used alternately with equal probability, the complexity of the proposed scheme per transmitted symbol is approximately half the complexity of the second DM. 

The reduced number of required DMs is also relevant with respect to problem~2. 
In fact, even if an efficient binary DM outputting bits with various probabilities is available \cite{Bohnke2020B}, our result indicates that the LLR can be quantized with negligible performance loss.  This is interesting from an implementation perspective.

Finally, since we consider sign-bit shaping, the stepwise MB-like distribution is implemented via only one bit level, whereas two levels are used in \cite{Gultekin2019}\cite{Iscan2018}. In light of problem 1, this is relevant when multi-level channel coding is used: In general, the last level does not need to be coded as the mutual information equals the entropy, $I(Y;B_m|B_1,...,B_{m-1}) = H(B_m|B_1,...,B_{m-1})$.
\section{Simplified sign-bit shaping}



%

\subsection{Description}

The Gaussian channel is expressed as 
\small
\begin{align}
Y=X+ Z, 
\end{align}
\normalsize
where $X$ is the channel input with alphabet $\mathcal{X}$ and $Z$ is the noise with distribution \small $\mathcal{N}(0, \sigma^2)$ \normalsize. 
The signal-to-noise ratio (SNR) is defined as \text{SNR} \small$= E[|X|^2]/E[|Z|^2]$\normalsize.

\begin{figure}[h]
\centering
\includegraphics[width=0.85\columnwidth]{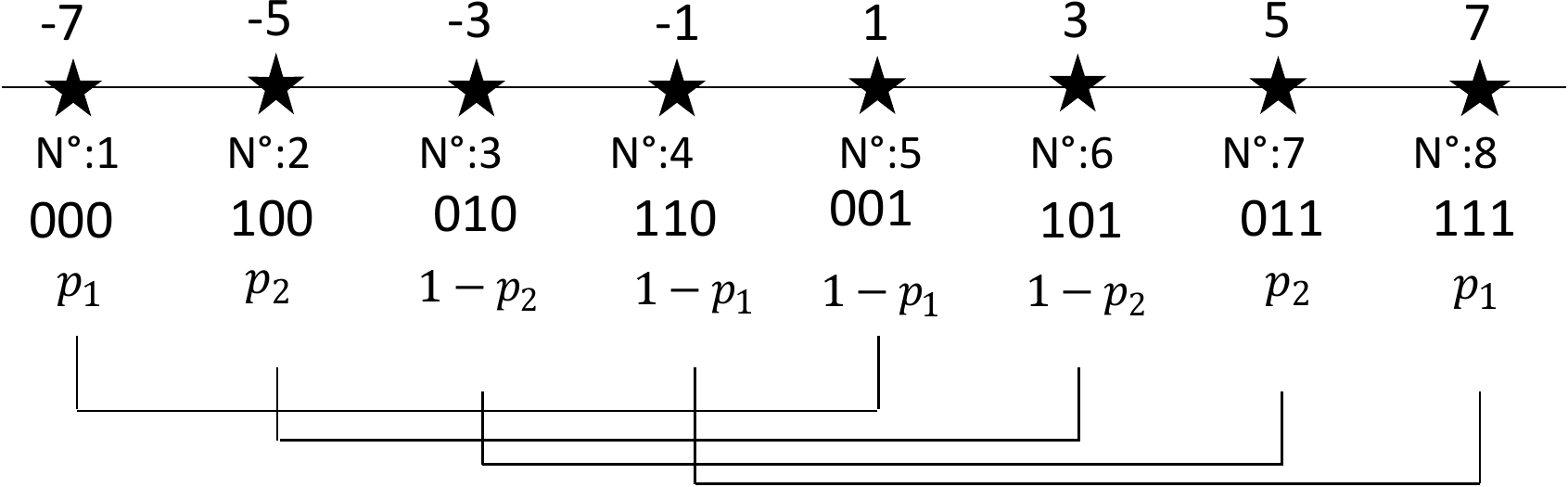}
\vspace{-2mm}
\caption{Illustration of the proposed distribution with the 8-ASK. 
The parameters $p_i$ represent the probabilities of the last bit level.
Natural labelling, with the less significant bits on the left, is also shown. }
\label{fig_Shaping_8PAM}
\end{figure}

Given a $M$-ASK constellation with natural labelling, we consider method 2 where only the last bit level is used for shaping. In other words, we consider sign-bit shaping. Consequently, the probability of each symbol is
\small
\begin{align}
\begin{split}
&\forall \ 1 \leq i \leq \frac{M}{2}, \  p(x_{i}) = p_i \cdot \left(\frac{1}{2} \right)^{m-1}, \\
&\forall \ \frac{M}{2} + 1 \leq i \leq M, \ p(x_{i}) = (1 - p_{i-\frac{M}{2}}) \cdot \left(\frac{1}{2} \right)^{m-1},
\end{split}
\end{align}
\normalsize
where the parameters $p_i = p(b_m|b_1,...,b_{m-1})$, $0\leq p_i \leq 1$, are to be optimized. 
Moreover, if the optimal distribution is symmetric, we set
\vspace{-2mm}
\begin{align}
\label{equ_proba}
p_i = 1 - p_{\frac{M}{2}-i+1}, 1\leq i \leq \frac{M}{4}.
\end{align}
An illustration is provided with the 8-ASK on Figure~\ref{fig_Shaping_8PAM}.
\vspace{-2mm}

Consequently, the MI should be optimized with respect to the set $\{p_i\}$, $1 \leq i \leq M/4$.
It can be expressed under the form of a convex optimization problem.
This is a simple extension of the fact that the channel capacity can be computed using convex optimization, see Problem 4.57 in \cite{BoydCVX}.
\vspace{-2mm}
\begin{align*}
\text{maximize} \ \  &I(X;Y), \\ 
\text{subject to} \  \ &p(x_i) =p_i \cdot \left( \frac{1}{2}\right)^{m-1} , \ 0 \le p_i \le 1,\\
                          & \sum_i p(x_i) \cdot x_i^2 \leq P.
\end{align*}
Of course, if the number of parameters $p_i$ is low, a brute-force line search can also be considered.

\begin{figure}[h]
\centering
\vspace{-3mm}
\includegraphics[width=0.95\columnwidth]{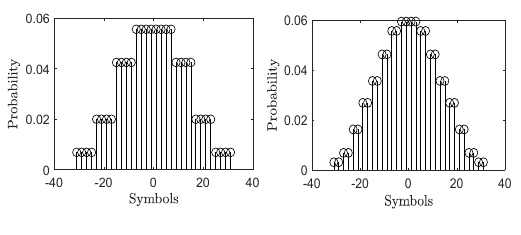}
\vspace{-7mm}
\caption{Optimized distribution at 24 dB for a 32-ASK for $\mathcal{P}=2$ and~$\mathcal{P}=4$.}
\vspace{-4mm}
\label{fig_distri_32ASK}
\end{figure}

As argued in the previous section, there is an interest to reduce the number of distinct $p_i$.
Consequently, we can force adjacent symbols to have the same probability. 
We let $\mathcal{P}$ be the number of distinct $p_i$.  
As an example, Figure~\ref{fig_distri_32ASK} shows how $\mathcal{P}$ changes the optimized distribution for the 32-ASK.

In Section~\ref{sec_perf}, we show that even for a $M$-ASK with large $M$, the performance remains satisfactory if $\mathcal{P}=2$. 


\subsection{Implementation of the proposed method}
\label{sec_implem}


\subsubsection{First implementation}

Let us define the function 
$D(b_1,b_2,...,b_{m})= \sum_{i=1}^{m}2^{i-1} \cdot b_i$,
which returns the decimal value of a binary sequence with natural labelling.
We assume that several binary sources $S_0$, $S_1$,...,$S_{M/4}$ are available, where $S_0$ generates  i.i.d. bits with a Bernoulli(1/2) distribution, and $S_i$ i.i.d. bits with a Bernoulli($p_i$) distribution.

\begin{figure}[h]
\centering
\includegraphics[width=1\columnwidth]{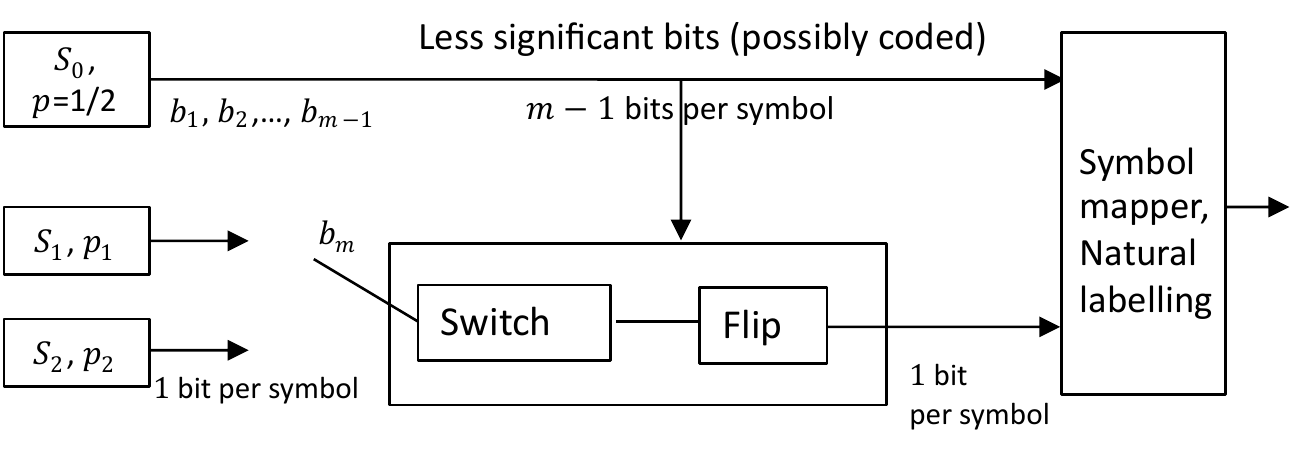}
\vspace{-5mm}
\caption{Shaping encoder with $\mathcal{P}=2$.}
\label{fig_system_nat_lab}
\end{figure}

With natural labelling (as on Figure~\ref{fig_Shaping_8PAM}), 
the shaping method can be implemented as illustrated on Figure~\ref{fig_system_nat_lab}:
First, the binary source $S_0$ generates $m-1$ bits $b_1,b_2,...,b_{m-1}$.  This corresponds to the bit levels that are not shaped. A switch then selects a source according to the required distribution $p_i=p(b_m|b_1,b_2,...,b_{m-1})$:
If $1\leq D(b_1,b_2,...,b_{m-1})+1 \leq M/4$, the switch selects the source $S_i$ of index $i=D(b_1,b_2,...,b_{m-1})+1$. One bit $b_m$ is then obtained from this source (and not flipped).  If $M/4+1\leq D(b_1,b_2,...,b_{m-1 }) +1 \leq M/2$, a switch selects the source $S_i$ of index $i=M/2-D(b_1,b_2,...,b_{m-1})$ and the bit $b_m$ obtained is then flipped\footnote{ 
 Consequently, at the receiver, the decoded shaping bit is flipped if $M/4+1 \leq D(\widehat{b_1},\widehat{b_2},…,\widehat{b_{m-1 } }) +1 \leq M/2$, where $\widehat{b_1},\widehat{b_2 },…,\widehat{b_{m-1}}$ represent the values of the decoded less significant bits.}. The flipping operation, which exploits the symmetry described by \eqref{equ_proba}, enables to divide by two the number of distinct binary sources (i.e., we generate a Bernoulli($1-p_i$) source from a Bernoulli($p_i$) source). 
Then, a symbol mapper generates the symbol to be transmitted based on the value of the $m$ bits.

If several adjacent symbols are constrained to have the same probability, i.e., $\mathcal{P}<M/4$ and groups of $M/(4\mathcal{P})$ symbols have the same probability, then the switch selects the first source for $1\leq D(b_1,b_2,...,b_{m-1} ) +1 \leq M/(4\mathcal{P})$, the second source for $1+ M/(4\mathcal{P}) \leq D(b_1,b_2,...,b_{m-1} ) +1 \leq 2 \cdot M/(4\mathcal{P})$, ..., and the $\mathcal{P}$-th source for  $1+ (M/4-M/(4\mathcal{P})) \leq D(b_1,b_2,...,b_{m-1} ) +1 \leq M/4$.
\vspace{-2mm}

\vspace{-2mm}
\subsubsection{Second implementation}

If only one binary source is available and the bits should be processed packetwise, the encoder can be adapted as follows. 
We consider the case $\mathcal{P}=2$. The extension to any $\mathcal{P}$ is straightforward.

First, $(m-1) k+n/2 (H(p_1)+H(p_2))$ bits are generated by the source $S_0$. $(m-1)k$ bits are encoded via an error-correcting code to yield $(m-1)n$ bits. In parallel, $n/2 \cdot H(p_1)$ bits and $n/2 \cdot  H(p_2)$ bits are processed by two binary DMs which output a sequence of $n/2$ bits where each bit is equal to 0 with a probability $p_1$ and $p_2$, respectively. Each of the $n$ sets of $m-1$ bits, at the output of the error-correcting code, controls the switch which selects one bit at the output of the corresponding DM. 

%

Of course, it is possible that the switch requests $n/2+\alpha$ bits from the first DM and $n/2-\alpha$ from the second DM, where $\alpha$ is a random quantity. To address this issue, we simply add the rule that if one source has no more bit available, the switch selects a bit from the other source. 
In Appendix~\ref{sec_app_switch}, we show that the impact of this rule on the performance is negligible.

Unlike with method 1, where all binary DMs in parallel are used for all transmitted symbols, the two binary DMs are used alternately with equal probability. This is an interesting property if one binary DM has a higher complexity than the other: The ``complex" binary DM is only used for one out of two transmitted symbols.

\section{Performance with ideal binary DMs}
\label{sec_perf}

\subsection{Mutual information}

\begin{figure}[h]
\centering
\vspace{-5mm}
\includegraphics[width=0.94\columnwidth]{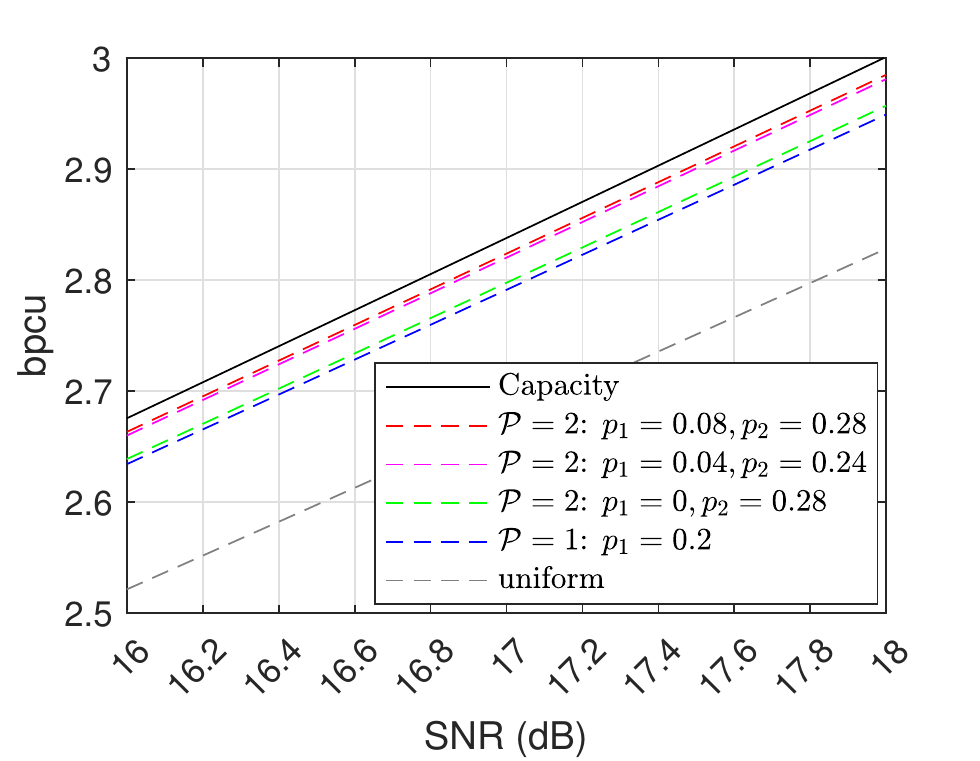}
\vspace{-3mm}
\caption{Impact of $\mathcal{P}$ and the $p_i$ on the MI with a 32-ASK. }

\label{fig_MI}
\end{figure}

We investigate the impact of $\mathcal{P}$ and the parameters $p_i$ on the performance for the 32-ASK constellation.
Figure~\ref{fig_MI} reports the MI obtained with a 32-ASK when optimizing the distribution with  $\mathcal{P}=1$ and $\mathcal{P}=2$ for several values of $p_1,p_2$.
With $\mathcal{P}=2$ and the optimal parameters $p_1=0.08,p_2=0.28$, the loss is less than 0.1 dB for information rates $R < 3$ bits per channel use (bpcu). Besides, the additional loss with $p_1=0.04,p_2=0.24$ is only 0.02 dB.
 Note that even $\mathcal{P}=1$  yields satisfactory performance.


For the 64-ASK (not on the figure), $\mathcal{P}=2$ yields a loss of approximately 0.2 dB at 5 bpcu.

\subsection{Block error rate}
\label{sec_block_error_rate}

\begin{figure}[t]
\centering
\vspace{-5mm}
\includegraphics[width=1\columnwidth]{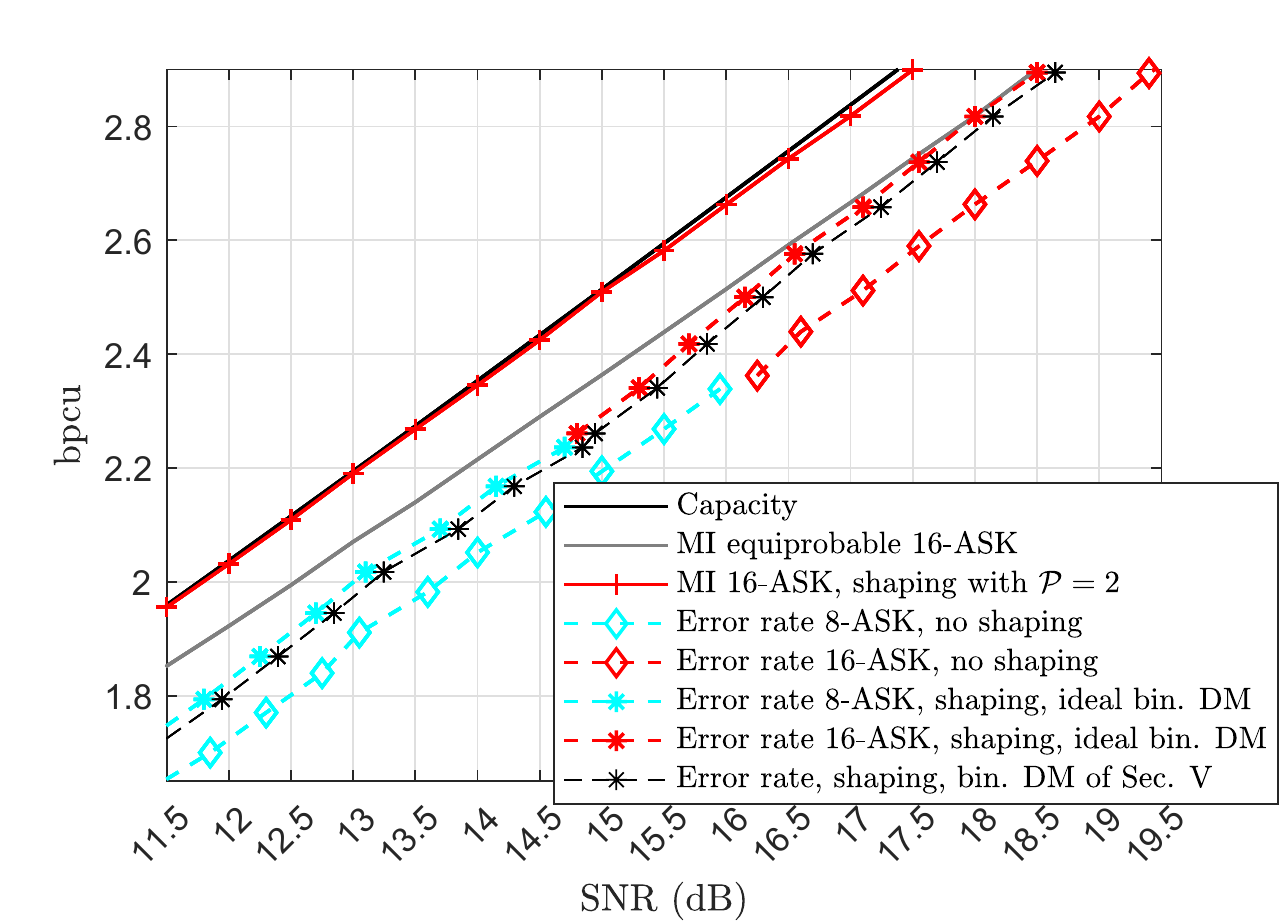} 
\vspace{-6mm}
\caption{Performance of the proposed shaping method, implemented via the system depicted on Figure~\ref{fig_system_nat_lab}, combined with multi-level polar coding.  
We also show the performance of the coded system without shaping for comparison. 
} 
\label{fig_main_perf}
\vspace{-5mm}
\end{figure}

 Figure~\ref{fig_main_perf} reports the performance  of the encoder of Figure~\ref{fig_system_nat_lab}, in terms of block error rate, where the less significant bits are encoded via multi-level polar coding with a block length $n=1024$.
Both the achievable rate without shaping and with shaping are shown. 
For the case with shaping, we show the performance assuming an ideal binary DM as well as the performance with the specific DM of the following section (see the end of Sec.~\ref{perf_with_DM}).
The scheme is evaluated for a block error rate of $10^{-2}$ via Monte Carlo simulations. 
We observe that the difference between the error-rate curves is similar to the difference between the MI curves with and without shaping. In other words, the shaping operation does not impact the performance of the error-correcting code.

\section{Performance with a specific binary DM}
\label{perf_with_DM}

We illustrate the benefit of having a small $p$ via an analysis of the algorithm of \cite{Schalkwijk1972} used as binary DM.
Similarly to arithmetic coding \cite{Ramabadran1990}, it enables to index binary sequences of length $n$ and weight $w=np$
via walking in Pascal's triangle.  We denote by $\mathcal{C}$ the set of all  $\binom{n}{w}$ such sequences. 
The difference between \cite{Schalkwijk1972} and \cite{Ramabadran1990} lies in the rule for walking in the triangle, where both walks start from the coefficient $\binom{n}{w}$.
We chose the algorithm \cite{Schalkwijk1972} because the complexity depends on $p$ in a more explicit manner than \cite{Ramabadran1990}. The latter algorithm could also benefit from a small $p$, but providing an explicit formula requires additional work and is left as future study.


It is proved in \cite{Schalkwijk1972} that a sequence $c=(c_1,...,c_n) \in \mathcal{C}$ can be indexed as
\begin{align}
i(c) = \sum_{k=1}^n c_k \binom{n-k}{w_k},
\end{align}
where $w_k$ is the weight of the sequence $(c_k,...,c_n)$.
Consequently, given an index $i(c)$, the corresponding sequence $c$ can be found via $w$ steps, where the two first steps are the following:
\begin{enumerate}
\item Find $1 \leq k_1 \leq n$ such that $\binom{k_1}{w}\le i(c) < \binom{k_1+1}{w}$. Set $c_1=0,...,c_{n-(k_1+1)} = 0$, and $c_{n-k_1} =1$.
\item Find $k_2 \leq k_1$ such that $\binom{k_2}{w-1} \le i(c) -  \binom{k_1}{w} < \binom{k_2+1}{w-1}$. Set $c_{n-(k_1-1)}=0,...,c_{n-(k_2+1)} = 0$, and $c_{n-k_2} =1$.
\end{enumerate}
We easily see that each step involves a binary search among at most $n$ elements. 
The worst-case complexity of the binary-search algorithm is log$_2(n)$. 
Consequently, the worst-case complexity of the algorithm per bit\footnote{Here, as in \cite{Pikus2017}, a comparison operation is counted as one operation for simplicity. However, a deeper analysis should take it into account since the magnitude of the numbers impacts the complexity (an issue also encountered with arithmetic coding). The memory aspect is also neglected.} is bounded from above by 
$p  \text{log}_2(n)$.
If two binary DMs of length $n/2$ are used in parallel as on Figure~\ref{fig_system_nat_lab}, with parameters $p_1=0.04$ and $p_2=0.24$, the complexity of the shaping scheme per transmitted symbol becomes $(p_1/2+p_2/2) \log_2(n/2) = 0.14 \log_2(n/2)$. 

For an output sequence of size $n$, the size $k$ of the equiprobable input sequence of the considered DM is the greater $k$ such that $2^k\leq \binom{n}{np}$.
The rate achieved (in bpcu) is therefore $\lfloor \log_2 \binom{n}{np} \rfloor / n$, to be compared with the optimal rate $H(p)$.
Hence, the rate loss is $H(p) - \lfloor \log_2 \binom{n}{np} \rfloor / n$.
Using the slope of the MI curves on Figure~\ref{fig_MI}, this yields an energy loss (which is approximately the same for all $p$) of about 0.04 dB for $n=800$, 0.03 dB for $n=1000$,  0.02 dB for $n=2000$, and 0.01 dB for $n=3400$.

For the overall shaping scheme, the rate loss with a block length $n$ is the sum of: The loss due to the quantization of the Gaussian, the average loss of the two binary DMs of length $n/2$ due to the finite length, and the loss due to the switch.
As an example, with the 32-ASK, two parameters $p_1=0.04$, $p_2=0.24$, and $n=2^{11}$, the rate loss with respect to the capacity at 17dB is $\approx 0.1+0.03 + 0.015 =  0.145$ dB. The performance curve with these parameters is shown on Figure~\ref{fig_main_perf}.

%

\vspace{-1mm}
\section{Conclusions}

In this paper, we proposed a simple sign-bit shaping scheme for high spectral-efficiency coding with natural labelling of the symbols.
We force several adjacent symbols to have the same probability, which results in a quantized version of
a MB-like distribution. 
Consequently, only two binary DMs are required, where one binary DM has a parameter $p$ close to 0.
This results in a new low-complexity shaping scheme at the cost of little performance loss.
\vspace{-4mm}
\section{Appendix}
\label{sec_app_switch}

%


The impact of the switch can be computed as follows. The real probabilities of the binary sources (with the switch) are
\small
\vspace{-1mm}
\begin{align}
\begin{split}
p_1'= \frac{(\frac{n}{2}-\epsilon) \cdot p_1 + \epsilon \cdot p_2}{n/2} = p_1 + \frac{2 \epsilon}{n} (p_2-p_1 ),\\
p_2'=  \frac{(\frac{n}{2}-\epsilon)\cdot p_2 + \epsilon \cdot p_1}{n/2} = p_2 +  \frac{2 \epsilon}{n} (p_1-p_2 ),
\end{split}
\end{align}
\normalsize
where $\epsilon$ is a constant obtained as follows. 
Let $X$ be a binomial distribution which assesses the probability to have $k$ success with $n$ Bernoulli(1/2) trials, i.e.,
\small
\vspace{-1mm}
\begin{align}
P(X=k) = {n \choose k} \left( \frac{1}{2}\right)^n.
\end{align}
\normalsize
The parameter $\epsilon$ is computed as follows: $\epsilon = E[k' \cdot X]$, where $k'=0$ if $k \le n/2+1$ and $k'= k - n/2$ if $k \geq n/2+1$ (where $k$ is the outcome of $X$), i.e., 

\small
\begin{align}
\epsilon = \sum_{k=\frac{n}{2}+1}^n (k-\frac{n}{2}) \times{n \choose k}\left(\frac{1}{2}\right)^n.
\end{align}
\normalsize

Let $E_s'$ be the energy of the constellation with the probabilities $p_i'$ and $E_s$ with the probabilities $p_i$. The energy loss due to the switch can be quantified as $\Delta = 10 \log_{10}(E_s'/E_s)$. For both the $8$-ASK and $32$-ASK with the parameters $p_1=0.04$ and $p_2=0.24$, we find $ \Delta= 0.01$ dB for $n=2^{12}$, $\Delta = 0.015$ dB for $n=2^{11}$, $\Delta = 0.02$ dB for $n=2^{10}$, $\Delta =0.03$ dB for $n=2^9$, and $\Delta =0.04$ dB for $n=2^8$.

\end{document}